\documentclass[twocolumn,prb,showpacs,floatfix]{revtex4}
\usepackage{graphicx}
\usepackage{dcolumn}
\usepackage{bm}
\usepackage{amsmath}

\begin{document}

\title{Edge states in graphene quantum dots: Fractional quantum Hall effect 
analogies and differences at zero magnetic field}

\author{Igor Romanovsky}
\author{Constantine Yannouleas}
%\email{Constantine.Yannouleas@physics.gatech.edu}
\author{Uzi Landman}
%\email{Uzi.Landman@physics.gatech.edu}

\affiliation{School of Physics, Georgia Institute of Technology,
             Atlanta, Georgia 30332-0430}

\date{17 December 2008}

\begin{abstract}
We investigate the way that 
the degenerate manifold of midgap edge states in quasicircular graphene 
quantum dots with zig-zag boundaries supports, under {\it free-magnetic-field\/} 
conditions, strongly correlated many-body behavior analogous to the fractional 
quantum Hall effect (FQHE), familiar from the case of semiconductor
heterostructures in high magnetic fields. Systematic exact-diagonalization (EXD)
numerical studies are presented for the first time for $ 5 \leq N \leq 8$ fully 
spin-polarized electrons and for total angular momenta in the range of
$ N(N-1)/2 \leq L \leq 150$. We present a derivation of a 
rotating-electron-molecule (REM) type wave function based on the 
methodology introduced earlier [C. Yannouleas and U. Landman, Phys. Rev. B  
{\bf 66}, 115315 (2002)] in the context of the FQHE in two-dimensional
semiconductor quantum dots. The EXD wave functions are compared with FQHE trial
functions of the Laughlin and the derived REM types. It is 
found that a variational extension of the REM offers a better 
description for all fractional fillings compared with that of the Laughlin 
functions (including total energies and overlaps), a fact that reflects the 
strong azimuthal localization of the edge electrons. In contrast with the 
multiring arrangements of electrons in circular semiconductor 
quantum dots, the graphene REMs exhibit in all instances a single $(0,N)$ 
polygonal-ring molecular (crystalline) structure, with all the electrons 
localized on the edge. Disruptions in the zig-zag 
boundary condition along the circular edge act effectively as impurities that
{\it pin\/} the electron molecule, yielding single-particle densities with
broken rotational symmetry that portray directly the azimuthal localization of 
the edge electrons.
\end{abstract}


\maketitle

\section{Introduction}

Since the discovery \cite{tsui82} of the fractional quantum Hall effect (FQHE) 
in two-dimensional (2D) semiconductor heterostructures in the presence of a high 
perpendicular magnetic field (${\cal B}$), phenomena associated with strongly 
correlated electrons in the lowest Landau level (LLL) have attracted 
significant and continuous attention. \cite{laug83,hald83,halp84,jain89,laug99,
jainbook,yann02,yann03,yann04,yann07,chan05,chan06}
Early on, it was realized that the essential 
many-body physics in the LLL could be most effectively grasped through the use 
of trial wave functions, with celebrated examples being the Jastrow-type 
Laughlin \cite{laug83} (JL) and composite fermion \cite{jain89} (CF) trial 
functions associated with the formation of a special class of quantum-liquid 
states. \cite{laug99} Later interest in finite 2D electronic systems, like 
semiconductor quantum dots (QDs) under high ${\cal B}$, 
led to the consideration of a 
different class of analytic trial functions known as rotating electron (or 
Wigner) molecules \cite{yann02,yann03,yann04,yann07} (REMs or RWMs). An 
advantage of the REMs is that, while they exhibit good total angular momenta, 
they directly incorporate the molecular (crystalline) configurations that 
dominate the anisotropic pair correlation functions revealed through numerical 
exact-diagonalization studies for a finite number of electrons under high 
${\cal B}$ in a disk geometry. The initial derivation \cite{yann02} of the REM 
trial functions generated a flare of theoretical 
activity around the question which class of 
trial functions (or combination of them) is most appropriate for describing the 
correlated many-body physics in the LLL of a small number of electrons $N$. 
\cite{jainbook,yann03,yann04,yann07,chan05,chan06} Furthermore, experimental 
advances in the field of ultracold trapped neutral atoms have been followed by 
considerable theoretical activity regarding the nature of correlated states in 
the LLL that are formed during the rapid rotation of the trap; see, 
e.g., Refs. \onlinecite{gunn00,gunn01,popp04,joli05,barb06,baks07}.

%% BEGIN INSERT %%%%%%%%%

Recent progress in the fabrication of new materials, and in particular in 
isolating and handling of a single graphene sheet, 
\cite{geim04,dehe04,geim05,stor05} offers most promising materials for future, 
post-silicone, miniaturized electronics\cite{geim07,dehe08} (sometime called 
nanoelectronics). This expectation is based on the two-dimensional character of 
graphene, where the electrons are essentially confined in the spatial direction
normal to the graphene plane. Fabrication of nanoscale device elements for use 
in electronics, spintronics and information processing, such as single-electron 
transistors, quantum point contacts, and quantum dots, would require additional
confinement in the other two spatial dimensions. However, to achieve the 
requested additional confinement, techniques 
(based on electrostatic gating) developed for 
the creation of QDs in semiconductors (such as GaAs) cannot be used because of 
the unique electronic structure of graphene. The difficulty originates from the 
relativistic, Dirac-like, nature of the low-energy quasiparticles in graphene. 
In particular, the gapless nature of the electrons in graphene\cite{guin08} 
allows them to penetrate unimpeded through a high and wide potential barrier. 
\cite{geim06} This phenomenon, which is known as the Klein paradox,
\cite{klei29,calo99} is in fact not a paradox but a consequence of the 
relativistic character of the electrons, with a sufficiently high potential 
being repulsive for electrons but attractive for positrons, thus resulting in 
positron states inside the barrier which can be matched to the electronic
continuum states outside, consequently resulting in perfect transmission through
the barrier; the underlying property of the  Dirac equation is known as the 
charge-conjugation symmetry.

In light of the above, one wishes to explore alternative, non-electrostatic 
methods, for fabrication of lower dimensionality graphene nanostructures. One 
route for achieving the desired added planar confinement consists of etching, or
cutting graphene into the desired geometry (e.g., ribbons, 
\cite{avou07,han07,dai08} circular disks, or other shapes
\cite{ihn08,ihn09,ihn09.2}). 
It is expected that further progress in fabrication, characterization and 
understanding of the properties of such graphene nanostructures (in particular 
zero-dimensional QDs) would lead to their use for the study of interesting 
many-body phenomena, as well as their employment as components in miniaturized 
electronic devices. 

Here we explore theoretically certain properties of circular graphene quantum 
dots, defined via cutting the desired shape from a two-dimensional extended 
sheet. In particular, we regard investigations of graphene QDs as providing an 
opportunity for reexamination (and possibly experimental resolution) of 
remaining questions concerning the appropriateness of liquid-type vs. 
molecular-type trial functions for a finite number of 2D electrons. Indeed, it 
has been known for some time that manifolds of degenerate midgap edge states 
exist in graphene nanostructures 
(such as graphene ribbons) when they terminate in a zig-zag boundary.
\cite{fuji96,dres96} In a recent paper \cite{wuns08} it was noted that the 
single-particle edge states associated with {\it circular\/} graphene dots with 
zig-zag boundary conditions, 
and in the absence of an applied magnetic field, display degeneracies and quantum
numbers in close analogy with the manifold of single-particle states that form 
the familiar LLL in semiconductor heterostructures at high ${\cal B}$. 
Furthermore, the 
numerical calculations of Ref.\ \onlinecite{wuns08}, covering rather limited 
ranges of electron numbers (i.e., $2 \leq N \leq 5$) and total angular momenta 
(i.e., $N(N-1) \leq L \leq 60$), suggested that the use of quantum-liquid-type 
trial functions in relation to the graphene LLL (gd-LLL) may be less promising 
than that of Wigner-crystal-type ansatzes.

In this paper, applying a methodology based on angular-momentum projection 
techniques that was introduced in Ref.\ \onlinecite{yann02}, 
we derive analytic REM trial functions appropriate for the 
gd-LLL. By introducing a single variational parameter, we demonstrate 
numerically (through systematic comparisons with EXD calculations) that
the variational variant of the REM (referred to as vREM) substantially 
outperforms the Laughlin trial functions (as well as the ansatz of Ref.\
\onlinecite{wuns08}) for all values of fractional fillings within the expanded
angular-momentum range $N(N-1)/2 \leq L \leq 150$, and for all of the
following larger numbers of electrons $N=5,6,7$ and 8.

\section{Lowest Landau level for circular graphene dots}

\subsection{Single-particle edge states}
\label{spedge}

It is well known \cite{ando05} that the low-energy bandstructure of graphene can
be described by a linearized tight-binding Hamiltonian $H$. For a graphene dot
with a circular symmetry this linearized Hamiltonian is given \cite{wuns08} 
in polar coordinates by
\begin{equation}
H = \hbar v_F \left(
\begin{array}{cc}
H_+ & 0  \\
0 & H_-
\end{array}
\right),
\label{tbham1}
\end{equation}
where
\begin{equation}
H_s = \left( 
\begin{array}{cc}
0 & e^{-is\phi} \left(-i \partial_r - \frac{s}{r} \partial_\phi \right) \\
e^{is\phi} \left(-i \partial_r + \frac{s}{r} \partial_\phi \right) & 0
\end{array}
\right),
\label{tbham}
\end{equation}
where $v_F$ is the Fermi velocity, and $s=\pm$ specifies the degenerate in
energy valleys for the two bands formed at the $K$ and $K^\prime$ points.
The index $s$ can be considered as a ``pseudospin'', which creates a fourfold
degeneracy when the spin degree of freedom is also considered.
The general solution of the Hamiltonian in Eq.\ (\ref{tbham}) is a two
component vector of the form
\begin{equation}
\left(
\begin{array}{c}
u_s^A (r,\phi)\\
u_s^B (r,\phi)
\end{array}
\right),
\label{genorb}
\end{equation}
where $A$ and $B$ denote the two triangular sublattices of graphene.

The usual volume solutions (which are zero on the graphene boundary but 
otherwise extend everywhere inside the area enclosed by the graphene dot) have 
energy $E_k=v_F k$, with $u_s^A$ and $u_s^B$ components that 
are expressed via the Bessel functions. Here we are not interested in 
such volume solutions. Instead we focus on the special edge states with zero 
energy $E=0$. These $E=0$ states are eigenfunctions of $H_s$ under the 
assumption that the graphene boundary exhibits an uninterrupted zigzag edge; 
\cite{fuji96,dres96,wuns08} an outline of their derivation from the Hamiltonian 
$H_s$ is given in Appendix A.

Henceforth we will only need to remember the precise form of the edge states,
which is given by 
\begin{eqnarray}
\left(
\begin{array}{c}
\psi_l^{A+}
\crcr
\psi_l^{B+}
\end{array}
\right)
=
\left(
\begin{array}{c}
\sqrt{\frac{l+1}{\pi R^{2(l+1)}}}
r^{l}  e^{il\phi}
\crcr 
0
\end{array}
\right)
\label{edge1}
\end{eqnarray}
and
\begin{eqnarray}
\left(
\begin{array}{c}
\psi_l^{A-}
\crcr
\psi_l^{B-}
\end{array}
\right)
=
\left(
\begin{array}{c}
0
\crcr
\sqrt{\frac{l+1}{\pi R^{2(l+1)}}}
r^{l}  e^{il\phi}
\end{array}
\right).
\label{edge2}
\end{eqnarray}

Namely one of the $A$ and $B$ components is everywhere zero (both on the
boundary and inside the dot) and the two valleys $+$ and $-$ are
decoupled even when the two-body Coulomb interaction is considered (which is
the main focus of this paper; see below).
As a result, in the following, we will drop the sublattice and
valley indices. We will also assume that the electrons are fully polarized.

Since the single-particle angular momentum $l \geq 0$ (to guarantee
normalizability), the manifold of such model edge states forms a set of
degenerate states similar to the lowest Landau level (LLL), familiar from
the case of 2D semiconductor devices at very high magnetic fields ${\cal B}$. 
We will call the manifold \cite{note2} of degenerate edge states with $l \geq 0$
the graphene-dot lowest Landau level 
(gd-LLL). The main difference [apart from the normalization constant, see
Eq.\ (3) in Ref.\ \onlinecite{yann02}] between the two cases is that the 
single-particle states in the usual LLL exhibit an additional Gaussian 
multiplicative factor $\exp(-r^2/4\Lambda_{\cal B}^2)$ where 
$\Lambda_B=\hbar c/(e {\cal B})$ is the magnetic length. This
Gaussian is missing from the expression for the edge states in Eqs.\
(\ref{edge1}) and (\ref{edge2}); instead one has $\psi_l \equiv 0$ for 
$r > R$. 

It is thus natural to investigate possible similarities related to fractional
quantum Hall effect (FQHE) physics.

\subsection{Classes of variational many-body wave functions}

FQHE physics in the LLL has been extensively investigated for the case of 2D 
semiconductor quantum dots.\cite{jainbook,yann07} A main 
focus has been the underlying nature of the correlated many-body states,
i.e., 'liquid' (Laughlin, composite fermions) or molecular ('crystalline', REM).
Detailed comparisons of pair-correlations functions between JL/CF and REM states
with EXD ones support the view that the molecular (localized electrons) picture 
in semiconductor QDs provides the most appropriate description. The emergence
of a gd-LLL, as described above in graphene dots offers a further area for
testing the appropriateness of liquid-type variational wave functions (JL/CF) 
versus those that describe REMs. 

First we will proceed with deriving a modified REM trial wave function that 
takes into consideration the differences between the single-particle states which
span the usual LLL (zero-node 2D-harmonic-oscillator states) and the gd-LLL 
(edge states).

\section{Derivation of variational REM trial wave functions for graphene dots}
\label{secrem}

\subsection{Intermediary parameter-free REM functions}

REM analytical wave functions in the LLL for electrons in two-dimensional
semiconductor quantum dots were derived earlier in Ref.\ \onlinecite{yann02}. 
The physics underlying such a derivation is based on the theory of symmetry 
breaking at the mean-field level and of subsequent symmetry restoration via
projection techniques. \cite{yann07,note1} In particular, this approach 
consists of two steps: 

(I) At the {\it first step\/}, one constructs a Slater determinant 
$\Psi^N(z_1,\ldots,z_N)$ out of displaced single-particle states 
$u(z_j,Z_{0,j})$, $j=1,\ldots,N$, that represent the 
electrons localized at the positions $Z_{0,j}$, with (omitting the particle 
indices) $z=x+iy=re^{i\phi}$ and $Z_0=X_0+iY_0=R_0e^{i\phi_0}$.
Note that necessarily all electrons are localized radially on the edge of the
graphene dot, so that $R_0=R$.

\begin{figure}[t]
%  \begin{center}
%\centering\includegraphics[width=7cm]{one-orbital-gdot_v2.eps}
\centering\includegraphics[width=7cm]{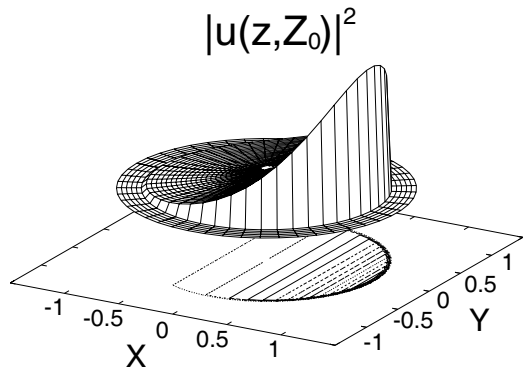}
%  \end{center}
\caption{%%%%
The displaced orbital $u(z,Z_0)$ (modulus square) representing a localized
electron at the point $Z_0=1+0i$. The radius $R$ of the dot serves as the unit
of length.
}
\label{disorbfig}% must come after the caption
\end{figure}

Naturally, for the LLL case of semiconductor QDs, the localized 
$u(z,Z_{0})$ single-particle states (referred to also as orbitals) were 
taken to be displaced Gaussians with appropriate Peierls phases due to the 
presence of a perpendicular magnetic field [see Eq.\ (1) in Ref.\ 
\onlinecite{yann02}]. In the case of electrons in graphene dots, 
however, the gd-LLL is spanned by edge-like orbitals (without a Gaussian
factor), i.e.,
\begin{equation}
\psi_l(z)=\sqrt{\frac{l+1}{\pi R^2}}\frac{z^l}{R^l},
\label{edge3}
\end{equation}
and as a result the appropriate localized orbitals are taken to have an
exponential form
\begin{eqnarray}
u(z,Z_0)=G \exp(z/Z_0),
\label{disorb}
\end{eqnarray}
with $G$ being the normalization constant (depending only on $R$).
The fact that $u(z,Z_0)$ in Eq.\ (\ref{disorb}) represents a localized electon 
is illustrated in Fig.\ \ref{disorbfig}.

%****************************** begin figure 1 **************************
%****************************** end figure 1 **************************

%And the basis zero energy modes are:

%\begin{eqnarray}
%\psi_l
%=\sqrt{\frac{l+1}{\pi R_0^2}}
%\left(\frac{r}{R_0}\right)^l e^{il\phi}
%\end{eqnarray}
%

%
%and, therefore,
%
%\begin{eqnarray}
%z^l=\psi_l R_{0}^{l+1}\sqrt{\frac{\pi}{l+1}}
%\end{eqnarray}

The localized orbital can be expanded in a series over the basis functions
in Eq.\ (\ref{edge3}) in the following way
\begin{equation}
u(z,Z_0)=
\sum\limits_{l=0}^{\infty}
C_l(Z_0)\psi_l(z),
\label{expan}
\end{equation}
with
\begin{equation}
C_l(Z_0) =
G \frac{ \sqrt{\pi} R^{l+1} }{ l!\sqrt{l+1} }\frac{1}{Z_0^l}.
\label{coeff}
\end{equation}
When constructing the many-body Slater determinant $\Psi^N [z]$, one considers 
$N$ orbitals $u(z_j,Z_{0,j})$ representing $N$ electrons on a ring of radius
$R$ (the radius of the graphene dot) forming a regular polygon, i.e.,
\begin{equation}
Z_{0,j}= R e^{2\pi i (1-j)/N},\;\; 1 \leq j \leq N.
\label{posit}
\end{equation} 

The single Slater determinant $\Psi^N [z]$ represents a {\it static\/} electron
(or Wigner) molecule (REM or RWM). Using Eq. (\ref{expan}), one finds the
following expansion (within a proportionality constant):
\begin{eqnarray}
\Psi^N[z]&=&
%\left(\frac{1}{\sqrt{\pi R_0^2}}\right)^N
\sum\limits_{l_1,\dots, l_N=0}^{\infty}
\frac{\sqrt{(l_1+1)\dots(l_N+1)}}{R^{l_1+\dots+l_N}} \nonumber \\
&& \hspace{-1.0cm} 
\times C_{l_1}(Z_{0,1})\cdots C_{l_N}(Z_{0,N}) D(l_1,l_2,\dots, l_N),
\label{exppsi}
\end{eqnarray}
where $D(l_1,l_2,\dots, l_N) \equiv \mbox{det}
[z_1^{l_1},z_2^{l_2},\ldots,z_N^{l_N}]$; the elements of the determinant are
the functions $z_i^{l_j}$ , with $z_1^{l_1},z_2^{l_2},\ldots,z_N^{l_N}$ being
the diagonal elements.

(II) {\it Second step:\/}
The Slater determinant $\Psi^N [z]$ breaks the rotational symmetry and thus
it is not an eigenstate of the total angular momentum $\hbar \hat{L} =
\hbar \sum_{j=1}^N \hat{l}_j$. However, one can restore \cite{yann02,yann07} the
rotational symmetry by applying onto $\Psi^N [z]$ the projection operator
\begin{equation}
{\cal P}_L \equiv \frac{1}{2\pi} \int_0^{2\pi} d\gamma 
e^{i\gamma(\hat{L} - L)},
\label{projl}
\end{equation}
where $\hbar L$ are the eigenvalues of the total angular momentum.

When applied onto $\Psi^N [z]$, the projection operator ${\cal P}_L$ acts as
a Kronecker delta: from the unrestricted sum in Eq.\ (\ref{exppsi}) it picks
up only those terms having a given total angular momentum $L$ (henceforth we
drop the constant prefactor $\hbar$ when referring to angular momenta).
As a result the projected wave function $\Phi^N_L ={\cal P}_L \Psi^N$ is
written as (within a proportionality constant)
\begin{eqnarray}
\Phi^N_L[z]=
\sum_{l_1,\dots,l_N}^{l_1+\dots+l_N=L}
\frac{D(l_1,\dots,l_N)}{l_1!\dots l_N!}
e^{i(\phi^0_1 l_1+\dots+ \phi^0_N l_N)},\;\;
\label{phi1}
\end{eqnarray}
with $\phi^0_j=2\pi(j-1)/N$.

We further observe that it is advantageous to rewrite Eq.\ (\ref{phi1}) by 
restricting the summation to the ordered arrangements 
$l_1 < l_2 < \ldots < l_N$, in which case we get
\begin{eqnarray}
\Phi^N_L[z]&=&
\sum_{0\leq l_1 < l_2 < \dots < l_N }^{l_1+l_2+\dots+l_N=L}
\frac{D(l_1,\dots,l_N)}{l_1!\dots l_N!} \nonumber \\
&& \hspace{-1.0cm} \times 
\mbox{det}[e^{i\phi^0_1 l_1}, e^{i\phi^0_2 l_2},\ldots,e^{i\phi^0_N l_N}]. 
\label{phi2}
\end{eqnarray}

The second determinant in Eq.\ (\ref{phi2}) can be shown \cite{mathematica} to
be equal (within a proportionality constant) to the following product of sine 
terms times a phase factor (independent of the individual $l_j$'s):  
\begin{equation}
e^{i\pi(N-1)L/N}
\prod_{1\leq j < k \leq N} \sin\left[\frac{\pi}{N}(l_j-l_k)\right].
\label{sines}
\end{equation}

Thus, the final result for the REM wave function is (within a proportionality
constant):
\begin{eqnarray}
\Phi^{\text{REM}}_{N,L}[z] &=&
\sum_{0\leq l_1 <  l_2 < \dots < l_N }^{l_1+l_2+\dots+l_N=L}
\frac{D(l_1,l_2,\dots,l_N)}{l_1!l_2!\dots l_N!} \nonumber \\
&& \times \prod_{1\leq j < k \leq N} \sin\left[\frac{\pi}{N}(l_j-l_k)\right].
\label{remwf}
\end{eqnarray}

\subsection{Introducing the variational parameter}

As described below, we found that the agreement between the REM in graphene dots
and the EXD solutions can be improved in a nontrivial way by introducing 
variational parameters. In particular, we found that consideration of a single 
variational parameter $\alpha$ serves our purpose remarkably well. Specifically, one replaces the prefactor
\begin{equation}
\frac{1}{l_1! l_2! \ldots l_N!}
\label{pref1}
\end{equation}
in Eq.\ (\ref{remwf}) by the following expression:
\begin{equation}
\frac{[(l_1+1) (l_2+1) \ldots (l_N+1)]^{(1-\alpha)/2}}
{(l_1! l_2! \ldots l_N!)^\alpha}.
\label{pref2}
\end{equation}
 
We call the $\alpha$-optimized wave functions the {\it variational\/} REM 
functions (denoted by vREM). When $\alpha=1$, the vREM coincides with the 
parameter-free REM expression. We note that a single-parameter variational 
crystal-type wave function, but with a different dependence on the parameter 
$\alpha$, has also been employed in Ref.\ \onlinecite{wuns08}. The present 
choice of variational parameter [see Eq.\ (\ref{pref2})] produces substantially
better results (see below). From a practical point of view, we note that
the crystal-type wave function proposed in Ref.\ 
\onlinecite{wuns08} does not contain a ``less-than'' ordered-arrangement 
restriction in the summation indices $l_1, \ldots , l_N$, and as a consequence 
it generates an exponentially larger number of expansion terms, thus
greatly inhibiting numerical evaluations for larger $N$ and $L$.

\section{Exact diagonalization and two-body Coulomb matrix elements}
\label{secexact}

For a circular graphene QD comprising $N$ electrons in the gd-LLL,
the many-body hamiltonian ${\cal H}$ comprises only the two-particle 
interelectron Coulomb repulsion, i.e.,
\begin{equation}
{\cal H}=
\sum_{i=1}^{N} \sum_{j>i}^{N} \frac{e^2}{\kappa r_{ij}}~,
\label{mbh}
\end{equation}
where $\kappa$ is the dielectric constant and $r_{ij}$ denotes
the relative distance between the $i$ and $j$ electrons.

The REM wave functions $\Phi^{\text{REM}}_{N,L}$ derived in the previous
section will be compared to the EXD ones $\Phi^{\text{EXD}}_{N,L}$ that are
solutions of the exact diagonalization of the hamiltonian (\ref{mbh}) in the 
many-body Hilbert space spanned by the Slater determinants
\begin{eqnarray}
\Psi_{L}^{I} [z] = \frac{1}{\sqrt{N!}} 
\left\vert
\begin{array}{ccc}
\psi_{l_1}(z_1) & \dots & \psi_{l_N}(z_1) \\
\vdots & \ddots & \vdots \\
\psi_{l_1}(z_N) & \dots & \psi_{l_N}(z_N) \\
\end{array}
\right\vert,
\label{detexd}
\end{eqnarray}
where the single particle functions $\psi_{l}(z)$ are given by the edge
states of Eq.\ (\ref{edge3}) and the index $I$ counts the arrangements
$0 \leq l_1 <l_2 < \ldots < l_N$ with $l_1 + l_2 + \ldots + l_N = L$.

Namely, $\Phi^{\text{EXD}}_{N,L}$ is written as 
\begin{equation}
\Phi^{\text{EXD}}_{N,L} [z] =
\sum_I C^I_L \Psi_{L}^{I} [z],
\label{phiexd}
\end{equation}
and the exact diagonalization of the many-body Schr\"{o}dinger equation
\begin{equation}
{\cal H} \Phi^{\text{EXD}}_{N,L} [z]=
E^{\text{EXD}}_{N,L} \Phi^{\text{EXD}}_{N,L}[z]
\label{mbsch}
\end{equation}
yields the coeffiecients $C^I_L$ and the EXD eigenenergies 
$E^{\text{EXD}}_{N,L}$.

The matrix elements $\langle \Psi_{L}^{I} | {\cal H} | \Psi_{L}^{J} \rangle$
between the basis determinants [see Eq.\ (\ref{detexd})] are calculated using
the Slater rules \cite{szabobook} and taking into account that, in the gd-LLL, 
the many-body hamiltonian has contributions from the Coulomb interaction only,
i.e., 
\begin{equation}
{\cal H} = \sum_{i<j} \frac{e^2}{\vert z_i - z_j \vert}.
\label{mbham}
\end{equation}
Naturally, one also needs the two-body matrix elements of the Coulomb
interaction in the basis formed out of the single-particle edge states.
These matrix elements are given through appropriate analytic expressions.
Indeed by defining 
\begin{eqnarray}
&& M(m,n,k)= \\
&& \hspace{-0.7cm} \int d z_1\!\int d z_2\  
\psi^{\dagger}_{m+k}(z_1) \psi^{\dagger}_{n-k}(z_2)
\frac{1}{\vert z_1-z_2\vert}
\psi_{m}(z_1) \psi_{n}(z_2), \nonumber
\label{clmem}
\end{eqnarray}
one finds
\begin{eqnarray}
&& \hspace{-0.5cm} M(m,n,k)=
\frac{1}{R}
\frac{2\pi {\cal C} \sqrt{\pi}}{(2m+2n+3)}
\frac{\Gamma\left(k+\frac{1}{2}\right)}{\Gamma(k+1)}
\\
&&\hspace{-0.5cm}
\times\left[
\frac
{\Gamma\left(n+1\right)}
{\Gamma\left(n+2\right)}
~_{3}F_{2}
\left(
1/2 ,\  k + 1/2 ,\  n+1\ ;\  1
\atop
k+1 ,\  n+2  
\right) \right. \nonumber\\
&&\hspace{-0.5cm}
 \left.
+\frac
{\Gamma\left(m+k+1\right)}
{\Gamma\left(m+k+2\right)}
~_{3}F_{2}
\left(
1/2 ,\  k + 1/2 ,\  m+k+1\ ;\  1
\atop
k+1 ,\  m+k+2  
\right)
\right], \nonumber
\label{clme}
\end{eqnarray}
where $_{3}F_{2}$ is the generalized hypergeometric function \cite{weinbook}
at the point $x=1$, $\Gamma$ is the Gamma function, and
\begin{eqnarray}
{\cal C}=\frac{\sqrt{(m+k+1)\ (n-k+1)\ (m+1)\ (n+1)}}{\pi^2}.
\label{clmec}
\end{eqnarray}

\section{Numerical Results}

\subsection{EXD total energies}
\label{exd_tot_ene}

%
%****************************** begin figure 2 **************************
\begin{figure}[t]
%\centering{\includegraphics[width=7.5cm]{all-exd-enrg-n5-n8_v2.eps}}
\centering{\includegraphics[width=7.5cm]{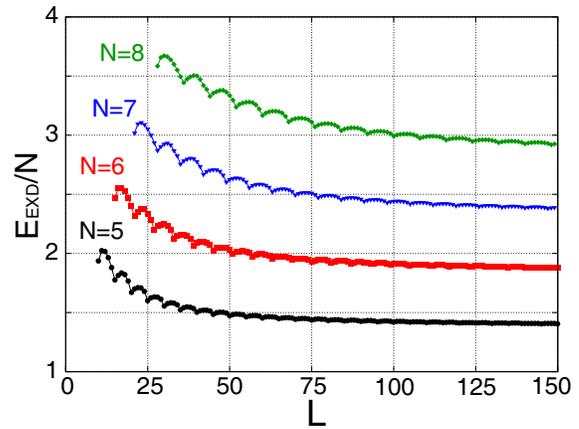}}
\caption{%%%%
Exact diagonalization ground-state energies in the graphene-dot LLL for $N=5$, 
6, 7, 8 electrons, as a function of the total angular momentum $L$.
Observe the appearance of cusp states of enhanced stability at the magic
angular momenta $L_m=N(N-1)/2+kN$, $k=0,1,2,\ldots$, a fact that indicates
formation of Wigner molecules having a single polygonal-ring configuration
$(0,N)$. Energies in units of $e^2/(\kappa R)$, with $\kappa$ being the
graphene dielectric constant and $R$ the radius of the quantum dot.
For $L \rightarrow \infty$, the ground-state energies approach asymptotically
the classical electrostatic energy [see Eq.\ (\ref{e_elst})].
}
\label{engs_n5_n8}
\end{figure}
%****************************** end figure 2 **************************

In Fig.\ \ref{engs_n5_n8} we display systematic EXD total energies in the range 
of $N=5$ to $N=8$ edge electrons as a function of the total angular momenta $L$
(in the large range $0 \leq L \leq 150$). This large $L$ range and the 
consideration of $N > 5$ electrons were not reached in another recent
publication;\cite{wuns08} they are, however, essential for unequivocally
establishing the proper similarities and differences with the 
high-magnetic-field physics of semiconductor QDs.

For fully polarized spins considered here, the minimum total angular momentum is
$L_0=N(N-1)/2$, in analogy with the case of semiconductor QDs.\cite{yann07} 
Furthermore, in analogy again with the case of semiconductor QDs, the total
energies decrease on the average as $L$ increases. On top of this average
trend, one observes prominent oscillations of period $N$. These oscillations
reveal that the states with $L=L_0+kN$, $k=0,1,\ldots$, are energetically
the most stable in their immediate neighborhood. Borrowing the terminology from
the literature \cite{yann07,maks00} of semiconductor QDs, we refer to these
states in graphene QDs as {\it cusp \/} states, and the corresponding total 
angular momenta (i.e., $L=L_0+kN$) as {\it magic angular momenta\/}. It is well 
known that cusp states develop to fractional-quantum-Hall-effect (FQHE) states in
the thermodynamic limit ($N \rightarrow \infty$), with the corresponding 
fractional filling factor being $\nu=L_0/L$.

Following a similar analysis\cite{yann07,maks00} with the case of semiconductor 
QDs, one can conclude that the appearance of the oscillatory period $N$ in the 
total energies (associated with the cusp states) is a reflection of formation of
$(0,N)$-type Wigner molecules, with all the electrons localized on a single
ring (of radius $R$) at the apices of a regular $N$-polygon. There is a major
difference, however, between the present system and the semiconductor quantum
dot case. That is, in semiconductor QDs, more than one isomers may form with 
concentric multiring arrangements occurring for $N > 5$ electrons in the dot; 
such arrangements 
are denoted as $(n_1,n_2,\ldots,n_q)$ (see Ref.\ \onlinecite{yann07}), where 
$n_r$, $r=1,2,\ldots,q$ are the number of localized electrons on each ring;
$\sum_{r=1}^q n_r=N$. In contrast, in the case of graphene dots 
only the one-ring $(0,N)$ molecular configuration arises (with no electron
residing at the geometrical center of the graphene QD).

For $L\rightarrow \infty$, the EXD energies in Fig.\ \ref{engs_n5_n8} approach 
the limiting value corresponding to the classical electrostatic energy of $N$ 
point-like electrons in a (0,N) configuration with radius $R$, i.e.,
\begin{equation}
E^{\text{cl}}(N)=
\frac{e^2}{4 \kappa R} N S_N,
\label{e_elst}
\end{equation}
with $S_N= \sum_{j=2}^{N} \left( \sin[(j-1)\pi /N] \right)^{-1}$.

%****************************** begin figure 3 **************************
\begin{figure}[t]
%\centering{\includegraphics[width=7.5cm]{exd-cpd-n7-l63_v2.eps}}
\centering{\includegraphics[width=7.5cm]{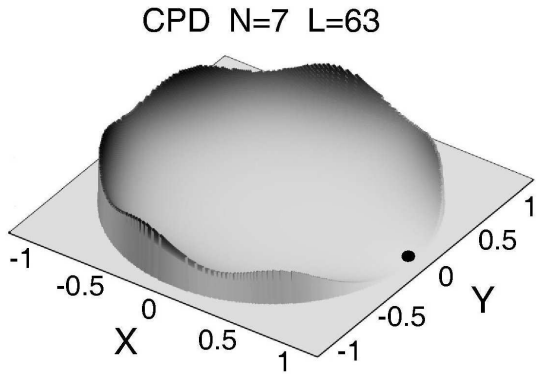}}
\caption{Conditional probability distribution [see Eq.\ (\ref{cpdexd})] 
associated with the EXD ground state
in the gd-LLL of $N=7$ electrons and for $L=63$ (corresponding to fractional
filling $\nu=1/3$). One clearly observes 6 humps in agreement with formation
of a $(0,7)$ Wigner molecule, and in contrast to the liquid-like Laughlin
physical picture. The fixed (observation) point is denoted by a solid dot. 
Lengths in units of the graphene-dot radius $R$. 
Vertical axis in arbitrary units.
}
\label{cpdn7l63}
\end{figure}
%****************************** end figure 3 **************************

\subsection{EXD densities and pair correlations}

The EXD eigenfunctions conserve the total angular momentum and the corresponding
electron densities are circularly symmetric. This property ``conceals'' the 
presence of the Wigner molecule. The crystalline structure, however, is present 
in the {\it intrinsic\/} frame of reference of the electron molecule, and it can
be revealed through the use of the fully anisotropic pair correlation
function $P(z,z_0)$, defined as
\begin{equation}
P(z, z_0)=
\langle \Phi^{\text{EXD}}_{N,L}| 
\sum_{i \neq j} \delta(z- z_i) \delta(z_0 - z_j)
|\Phi^{\text{EXD}}_{N,L} \rangle.
\label{cpdexd}
\end{equation}

$P(z,z_0)$ is often referred to as the conditional probability distribution
(CPD), since it is proportional to the probability of finding an electron at 
$z$ under the condition that another one is situated at the point $z_0$ 
(the socalled fixed, or observation, point).

In Fig.\ \ref{cpdn7l63} we display the conditional probability distribution
for the case of $N=7$ electrons and the magic total angular momentum $L=63$
($L-L_0=7k$, with $k=6$), which corresponds to the celebrated $\nu=1/3$
fractional filling. One clearly observes six humps (arranged in a single-ring
configuration) associated with formation of a (0,7) rotating Wigner molecule. 
(As is well known from the literature of semiconductor quantum dots,
\cite{yann07} the localized electron at the fixed point does not contribute any 
hump in the CPDs.) Similar CPDs are found for other values of $N$.

\subsection{Comparison between vREM and EXD wave functions}
\label{vREMcom}

%****************************** begin figure 4 **************************
\begin{figure}[t]
%\centering{\includegraphics[width=8.0cm]{all-exd-err-aa-n5-n8_v2.eps}}
\centering{\includegraphics[width=8.0cm]{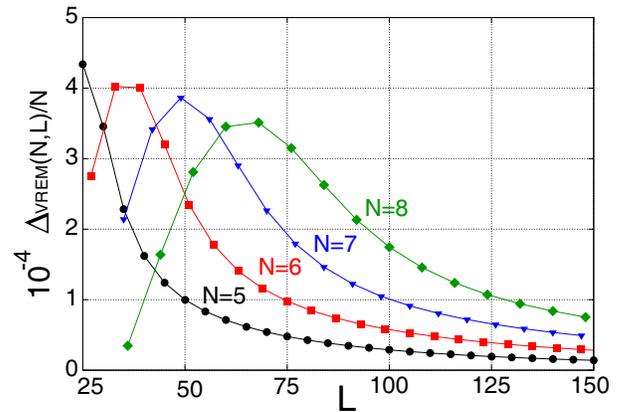}}
\caption{Relative error (per electron) of the vREM ground-state energies as a 
function of the total angular momentum $L$.
}
\label{erraa_n5_n8}
\end{figure}
%****************************** end figure 4 **************************

We turn now to comparisons between the EXD wave functions and the vREM ones.
We first observe that the REM and vREM functions [see Sect. \ref{secrem}] 
correspond to the magic angular momenta $L=L_0+k N$, $k=0,1,2,\ldots$, since all
the sine-product coefficients in the expansion (\ref{remwf}) are identically 
zero for $L \neq L_0+k N$. In this section, we will show that the vREM functions 
represent a high-quality approximation to the EXD eigenfunctions by investigating
wave function overlaps and relative errors between the total energies obtained 
by the two methods; the relative errors are defined as
$\Delta_{\text{vREM}} (N,L) = (E^{\text{vREM}}_{N,L}
- E^{\text{EXD}}_{N,L})/E^{\text{EXD}}_{N,L}$.

%****************************** begin figure 5 **************************
\begin{figure}[t]
%\centering{\includegraphics[width=8.0cm]{all-exd-ovl-aa-n5-n8_v2.eps}}
\centering{\includegraphics[width=8.0cm]{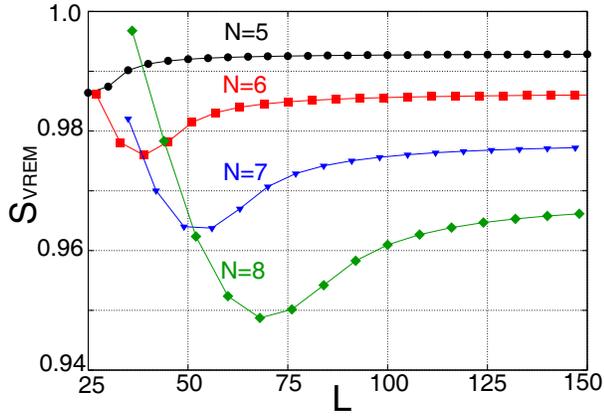}}
\caption{Overlaps of the vREM ground states with the EXD ones as a function of 
the total angular momentum $L$.
}
\label{ovlaa_n5_n8}
\end{figure}
%****************************** end figure 5 **************************

We start by displaying in Fig.\ \ref{erraa_n5_n8} the relative error 
$\Delta_{\text{vREM}} (N,L) $ of the vREM total energies. 
The vREM offers an excellent approximation, since the maximum relative
error is less than 0.045\%. For all sizes examined, the maximum relative error 
occurs about $\nu=1/3$ (see Section \ref{exd_tot_ene}), and subsequently it 
decreases as $L$ increases, approaching zero as $L \rightarrow \infty$.

In Fig.\ \ref{ovlaa_n5_n8}, we display the overlaps ${\cal S}_{\text{vREM}} 
\equiv \langle \Phi^{\text{EXD}}_{N,L} | \Phi^{\text{vREM}}_{N,L} \rangle $ 
between the vREM functions and the EXD solutions. These overlaps are larger than
0.985 for $N=5$ and larger than 0.95 for $N=8$ and they tend to slowly approach 
unity as $L$ increases. 

%****************************** begin figure 6 **************************
\begin{figure}[b]
%\centering{\includegraphics[width=8.0cm]{all-exd-alf-aa-n5-n8_v2.eps}}
\centering{\includegraphics[width=8.0cm]{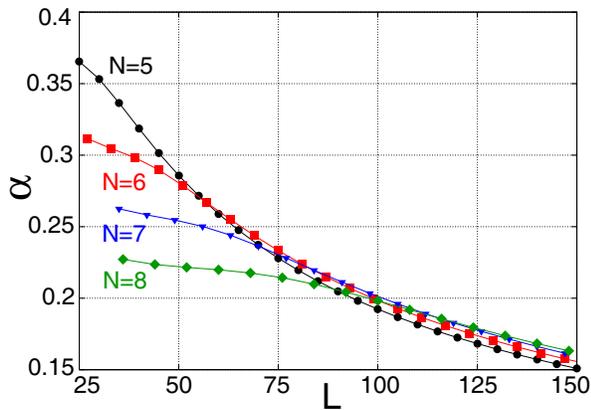}}
\caption{The values of the variational parameter $\alpha$ [see
Eq.\ (\ref{pref2})] that optimize the vREM trial functions for a given
$N$ as a function of $L$.  
}
\label{aa_n5_n8}
\end{figure}
%****************************** end figure 6 **************************

In Fig.\ \ref{aa_n5_n8}, we display the values of the variational parameter
$\alpha$ that optimize the vREM trial functions for a given number of electrons
$N$ as a function of $L$. These values are significantly different from
unity (which corresponds to the parameter-free REM). In fact the optimal
$\alpha$ values are smaller than 0.4, and they slowly decrease to about 0.16
for $L=150$, for all the values $5 \leq N \leq 8$. 
We stress that optimization of $\alpha$ is essential for
achieving a high quality reproduction of the EXD ground states. Without
the additional optimization (i.e., taking only the value $\alpha=1$) the behavior
of the overlaps is unsatisfactory, since they tend to diminish as $L$ increases 
(see Appendix B). The degradation of the overlaps of the parameter-free REM 
($\alpha=1$) as $L$ increases in the case of the graphene quantum dot 
contrasts with the opposite behavior of the overlaps of the parameter-free REM 
in the case of semiconductor quantum dots. \cite{yann02,yann03,yann07} 
This difference is attributed to the absence of translational invariance for the
electrons in the graphene quantum dot, which leads to differences in the 
organization of the EXD excitation spectra.

\subsection{EXD versus Laughlin wave functions}

It is interesting to compare the accuracy with which the vREM 
wave functions approximate the EXD ones with that of the Laughlin trial
functions. The Laughlin wave functions are restricted to the socalled
main (odd) fractions $\nu = 1/(2m+1)$ and have played an important role in
the FQHE literature of the extended two-dimensional electron gas in 
semiconductor heterostructures. Their form is
\begin{equation}
\Phi^{\text{Lauglin}}_{N,L} [z]=
\prod_{1 \leq i<j \leq N} (z_i-z_j)^{2m+1},
\label{wflau}
\end{equation}
where the Gaussian factors are missing [see Sect. \ref{spedge}] due to the 
differences in the single-particle states between semiconductor and graphene 
quantum dots; $m=0,1,2,\ldots$.

%****************************** begin figure 7 **************************
\begin{figure}[t]
%\centering{\includegraphics[width=8.0cm]{all-exd-err-lau-n5-n8_v2.eps}}
\centering{\includegraphics[width=8.0cm]{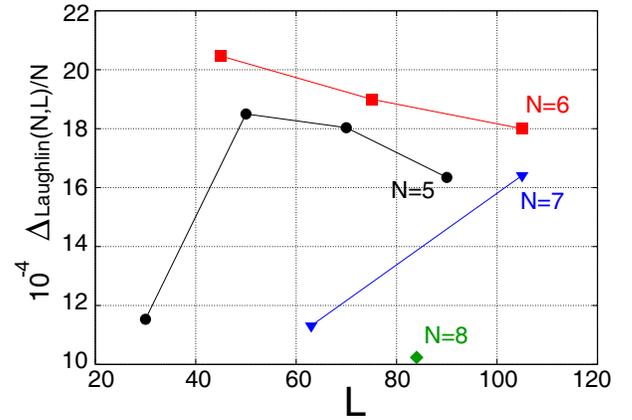}}
\caption{Relative error (per electron) of the Laughlin energies as a function of
the total angular momentum $L$.
}
\label{errlau_n5_n8}
\end{figure}
%****************************** end figure 7 **************************

%****************************** begin figure 8 **************************
\begin{figure}[b]
%\centering{\includegraphics[width=8.0cm]{all-exd-ovl-lau-n5-n8_v2.eps}}
\centering{\includegraphics[width=8.0cm]{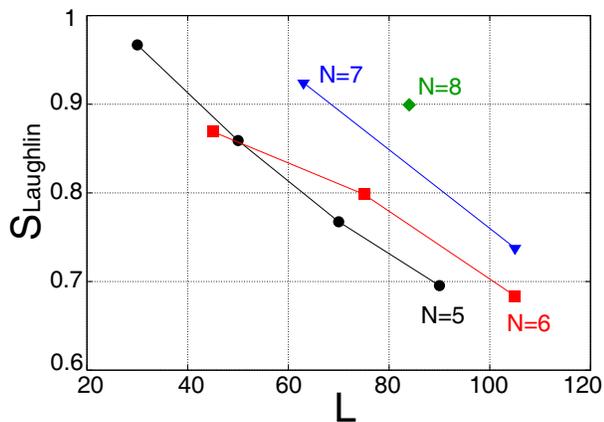}}
\caption{
Overlaps of the Laughlin trial states with the EXD ones as a function of
total angular momentum $L$.
}
\label{ovllau_n5_n8}
\end{figure}
%****************************** end figure 8 **************************

In Fig.\ (\ref{errlau_n5_n8}), we display the relative error,
$\Delta_{\text{Laughlin}} (N,L) = (E^{\text{Laughlin}}_{N,L}
- E^{\text{EXD}}_{N,L})/E^{\text{EXD}}_{N,L}$, of the Laughlin
total energies with respect to the ground-state EXD ones as a function of 
increasing total angular momentum $L$. The Laughlin relative errors are 
substantially larger (on the average by a factor of 5) than the vREM ones 
[see Fig.\ \ref{erraa_n5_n8}]; this is the case even for the celebrated 
$\nu=1/3$ fractional filling.

In addition, the Laughlin overlaps ${\cal S}_{\text{Laughlin}} \equiv
\langle \Phi^{\text{EXD}}_{N,L} | \Phi^{\text{Laughlin}}_{N,L} \rangle $
(plotted in Fig.\ \ref{ovllau_n5_n8}) exhibit an unsatisfactory performance 
compared to that of the vREM overlaps, that is: (i) they become steadily smaller
as the angular momentum increases, and (ii) even for $\nu=1/3$, they are smaller 
than the corresponding vREM overlaps in all instances studied here, i.e., 
$N=5-8$ electrons in the graphene dot.

We conclude that the Laughlin functions fail to capture the case of the gd-LLL,
while the vREM functions offer an appropriate approximation for graphene QDs.

\subsection{EXD versus composite-fermion wave functions}

It is also interesting to compare the accuracy with which the vREM
wave functions approximate the EXD ones with that of the composite-fermion trial
functions, which are more general than the Laughlin functions. Along with the 
Laughlin functions, they CF trial functions have played a significant role in
the FQHE literature of the extended two-dimensional electron gas in
semiconductor heterostructures. Their form \cite{jain89,jainbook} in
the disc geometry (case of 2D QDs studied here) are given by the expression,
\begin{equation}
\Phi^{\text{CF}}_L (N) = 
{\cal P_{\text{LLL}}} \prod_{1 \leq i < j \leq N} (z_i - z_j)^{2m} 
\Psi^{\text{IPM}}_{L^*},   
\label{cfeqs}
\end{equation}
where $z=x+i y$ and $\Psi^{\text{IPM}}_{L^*}$ is the Slater determinant 
of $N$ {\it non-interacting\/} electrons of total angular momentum $L^*$; it
is constructed according to the Independent Particle Model (IPM) from the
Darwin-Fock\cite{df} orbitals $\psi^{\text{DF}}_{p,l}(z)$, where $p$ and $l$ are 
the number of nodes and the angular momentum, respectively [for the values of 
$p$ and $l$ in the $n$th Landau level in high ${\cal B}$, see Appendix F].

The single-particle electronic orbitals in the Slater determinant 
$\Psi^{\text{IPM}}_{L^*}$ are not restricted to the lowest Landau level.
As a result, it is necessary to apply a projection operator 
${\cal P_{\text{LLL}}}$ to guarantee that the CF wave function lies in the LLL, 
as appropriate for $B \rightarrow \infty$. We carry the ${\cal P_{\text{LLL}}}$
projection according to section 4.3 of Ref.\ \onlinecite{heinbook}. After
obtaining the projected CF function in the LLL, the corresponding trial function
in the gd-LLL is constructed by simply replacing $\psi^{\text{DF}}_{0,l}(z)$
by the $\psi_{l}(z)$ in Eq.\ (\ref{edge3}).  

Since the CF wave function is an homogeneous polynomial in the electronic
positions $z_j$'s, its angular momentum $L$ is related to the non-interacting
total angular momentum $L^*$ as follows,
\begin{equation}
L = L^* + m N(N-1)=2m L_0.
\label{ltol*}
\end{equation}

Here we will consider the mean-field version of the composite-fermion theory,
according to which the Slater determinants $\Psi^{\text{IPM}}_{L^*}$ are 
the socalled compact states (see Appendix F for details; the corresponding
values of $L^*$ are listed in Table \ref{cftapp}). 
We note that recently several extensions
of the CF theory have been formulated \cite{jeon07} that account for 
residual-interaction effects among the individual $\Phi^{\text{CF}}_L (N)$ 
composite-fermion states. Consideration of such residual-interaction effects is 
beyond the scope of the present paper.

In Table \ref{cftmain}, we compare total CF and EXD energies (per particle) for 
$N=6$ electrons in a graphene dot. We also display the corresponding relative 
errors $\Delta_{\text{CF}} (N,L)/N = (E^{\text{CF}}_{N,L}
- E^{\text{EXD}}_{N,L})/(N E^{\text{EXD}}_{N,L})$ and overlaps
${\cal S}_{\text{CF}} \equiv
\langle \Phi^{\text{EXD}}_{N,L} | \Phi^{\text{CF}}_{N,L} \rangle$.

%******************************** table I ********************
\begin{table}[t] %[H] add [H] placement to break table across pages
\caption{\label{cftmain}%
Total CF and EXD energies [per particle, in units of $e^2/(\kappa R)$] for $N=6$
electrons in a graphene dot. The corresponding relative errors, 
$\Delta_{\text{CF}}/N$, and overlaps, ${\cal S}_{\text{CF}}$, are also listed. 
For the determination of the auxiliary angular momenta $L^*$, see Appendix F.}
\begin{ruledtabular}
\begin{tabular}{lcccc}
$L(L^*)$ & $E^{\text{EXD}}/N$ & $E^{\text{CF}}/N$ & 
$10^{-4} \; \Delta_{\text{CF}}/N$  &  ${\cal S}_{\text{CF}}$ \\ \hline 
21(-9) &   2.31357  &    2.3548  &    29.67  &        0.887\\
51(-9) &   1.99693  &    2.0385  &    34.67  &        0.793\\
30(0)  &   2.24863  &    2.3473  &    73.17  &        0.369\\
60(0)  &   1.99452  &    2.0520  &    48.00  &        0.507\\
35(5)  &   2.15628  &    2.3022  &    112.8  &        0.356\\
65(5)  &   1.97413  &    2.0410  &    56.50  &        0.451\\
39(9)  &   2.06465  &    2.0952  &    24.67  &        0.892\\
69(9)  &   1.94477  &    1.9713  &    22.67  &        0.754
\end{tabular}
\end{ruledtabular}
\end{table}
%***************************** end table I ********************

In addition to the $L=15+6k$, $k=0,1,2,\ldots$ magic angular momenta for $N=6$ 
found from EXD calculations (see Fig.\ \ref{engs_n5_n8}), the compact-state CF 
theory mistakenly predicts the existence of magic angular momenta with 
$L=15+5k$, $k=0,1,2,\ldots$, e.g., for $L=30,35,60,65$. Furthermore, even for the
states with $L=15+6k$, e.g., $L=21,39,51,69$ (Table \ref{cftmain}), 
the quantitative 
performance of the compact CF functions (concerning relative errors and
overlaps) is rather weak compared to that of the vREM: the CF relative errors 
are larger roughly by a factor of 10, while the CF overlaps are systematically 
smaller ($<0.9$) than the vREM ones ($>0.97$) 
(see Figs.\ \ref{engs_n5_n8} and \ref{ovlaa_n5_n8}, and Table \ref{cftmain}). 

As was the case with the Laughlin functions, we conclude that the compact CF 
functions are also at a disadvantage compared to the vREM concerning the
description of strongly correlated states in the gd-LLL.

\subsection{Comparison with the Wigner-crystal ansatz of Ref.\ 
\onlinecite{wuns08}}

Here we compare the vREM total energies with those associated with the
Wigner-crystal trial function of Ref.\ \onlinecite{wuns08}, given by
\begin{eqnarray}
\Phi^{\text{WC}}_{N,L}[z] &=&
\sum_{l_1, l_2 , \dots , l_N }^{l_1+l_2+\dots+l_N=L}
\exp \left( {-i\sum_n \frac{2\pi n l_n}{N}} \right) \nonumber \\
&& \hspace{-0.5cm} \times \prod_n (l_n+1)^{w+1/2}  D(l_1,l_2,\dots,l_N),
\label{wcwf}
\end{eqnarray}
where $w$ is a variational parameter.

\begin{figure}[t]
%\centering{\includegraphics[width=8.0cm]{all-exd-err-ag-n5-n8_v2.eps}}
\centering{\includegraphics[width=8.0cm]{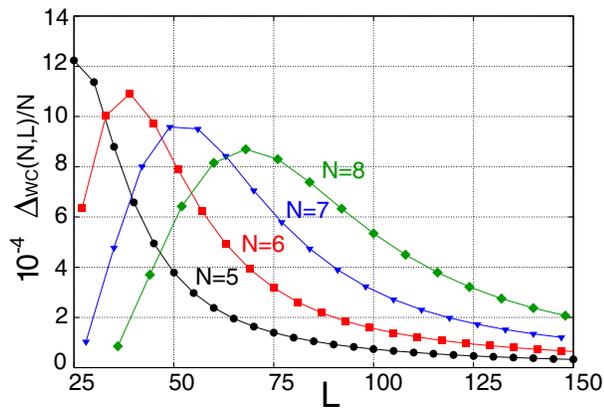}}
\caption{Relative errors (per electron) of the energies of the WC-ansatz [Eq.\ 
(\ref{wcwf})] as a function of total angular momentum $L$.
}
\label{errag_n5_n8}
\end{figure}

In Fig.\ \ref{errag_n5_n8}, we display the relative errors 
$\Delta_{\text{WC}} (N,L) = (E^{\text{WC}}_{N,L} - 
E^{\text{EXD}}_{N,L})/E^{\text{EXD}}_{N,L}$
of the WC-ansatz energies relative to the EXD ones. 
From a comparison of these results with those
displayed in Fig.\ \ref{erraa_n5_n8}, we conclude  
that the relative errors of the WC ansatz are on the average at least 
twice as large as those corresponding to the vREM, reflecting the superior
description of the gd-LLL provided by the latter function.

%****************************** begin figure 9 **************************
%****************************** end figure 9 **************************

\section{Pinned electron molecules}
\label{pinem}

The zig-zag geometry (on which the boundary conditions are applied; see Section
\ref{spedge}) does not allow formation of a continuous circular edge
without some structural or chemical modification of the graphene hexagonal
lattice structure. Without such modification, regions along the circular edge 
satisfying a zig-zag condition must necessarily be disrupted by a number of
discrete points associated with arm-chaired conditions.\cite{been08} 
It has been found that the edge states are robust in this case, 
\cite{dres96,been08} and as a result these discrete set of disruptions act as 
effective impurities that modify the many-body hamiltonian in Eq.\ 
(\ref{mbh}). The presence of such impurity terms in the many-body hamiltonian 
will mix the good-total-angular-momentum REM states, resulting in the formation 
of {\it pinned\/} electron molecules (PEMs). In contrast to the REMs (whose
electron density is uniform along the azimuthal direction, that is, not
showing any azimuthal density modulation), the electron density
of a pinned electron molecule is expected not to have circular symmetry; it will 
exhibit angular density oscillations, and the number of humps will equal the 
number of electrons $N$.

%****************************** begin figure 10 **************************
\begin{figure}[t]
%\centering{\includegraphics[width=8.0cm]{cmb_density_s1_n7_l35_l42_xy_v2.eps}}
\centering{\includegraphics[width=8.0cm]{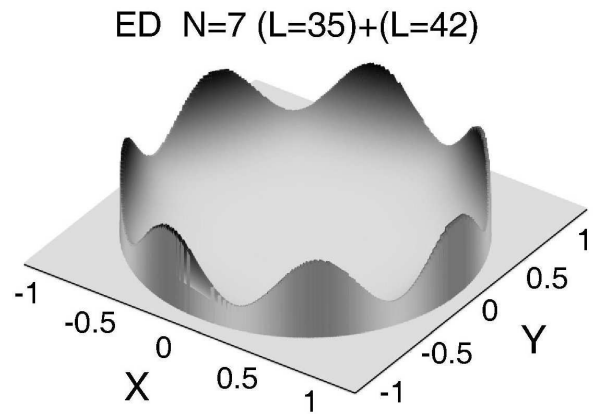}}
\caption{Electron density of a pinned molecule for $N=7$ electrons formed
from the linear superposition of two REM states with $L=35$ and $L=42$.
Lengths in units of the graphene dot radius $R$. Electron density in
units of $R^{-2}$.
}
\label{pemn7}
\end{figure}
%****************************** end figure 10 **************************

%****************************** begin figure 11 **************************
\begin{figure}[b]
%\centering{\includegraphics[width=8.0cm]{cmb_density_s1_n8_l44_l52_xy_v2.eps}}
\centering{\includegraphics[width=8.0cm]{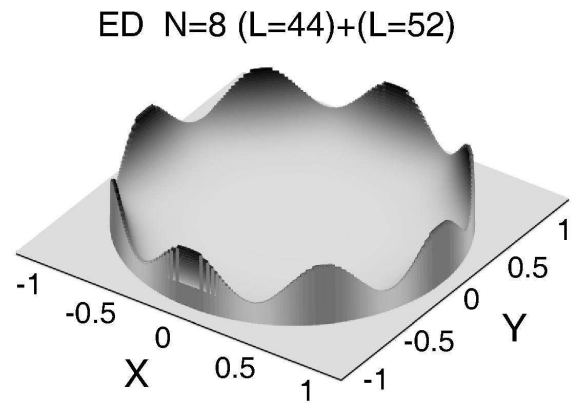}}
\caption{Electron density of a pinned molecule for $N=8$ electrons formed
from the linear superposition of two REM states with $L=44$ and $L=52$.
Lengths in units of the graphene dot radius $R$. Electron density in
units of $R^{-2}$.
}
\label{pemn8}
\end{figure}
%****************************** end figure 11 **************************

We demonstrate this property of a PEM for two particular cases displayed in 
Figs.\ \ref{pemn7} and \ref{pemn8}. Fig.\ \ref{pemn7} displays for $N=7$ the 
electron density for the linear superposition of two REM states with $L=35$ and
$L=42$, while Fig.\ \ref{pemn8} displays for $N=8$ the electron density for the
linear superposition of two REM states with $L=44$ and $L=52$.
In both cases the expected angular modulation is clearly well formed with seven
humps in the former and eight humps in the latter case. 

\section{Summary}

The manifold of degenerate midgap (zero-energy) edge states in circular graphene
quantum dots with zig-zag boundaries resembles, 
{\it under free-field conditions\/}, \cite{note3}
the celebrated lowest Landau level, familiar from the case of semiconductor
heterostructures in high magnetic fields. The effect of $e-e$ interactions
in this graphene-LLL were systematically investigated and were found to
generate many-body strongly correlated behavior that exhibits many similarities 
with the fractional quantum Hall effect. 

Numerical exact-diagonalization studies were presented for $ 5 \leq N \leq 8$ 
fully spin-polarized electrons and for total angular momenta in the range of
$ N(N-1)/2 \leq L \leq 150$. Moreover, we presented a derivation of a
rotating-electron-molecule type wave function based on the methodology 
introduced earlier\cite{yann02} in the context of the FQHE in two-dimensional
semiconductor quantum dots. The EXD wave functions were compared with 
the derived rotating-electron-molecule and other suggested FQHE 
trial functions, like the Laughlin function and the Wigner-crystal ansatz
of Ref.\ \onlinecite{wuns08}. It was
found that a variational extension of the REM offers a better
description for all fractional fillings compared with that of the Laughlin
and Wigner-crystal ansatz functions (including total energies and overlaps). 
The success of the REM function reflects the importance of strong azimuthal 
localization of the edge electrons in graphene quantum dots.

The variational REM functions were derived through the use of a two-step method:
 (i) first a mean-field-type single Slater determinant constructed out of $N$ 
localized electron orbitals (that break circular symmetry) was considered; this 
determinant describes the finite analog of a classical static Wigner-crystal, 
and (ii) a multideterminantal wave function was generated through 
the subsequent application of projection techniques that introduced azimuthal 
fluctuations and restored the circular symmetry and good total angular momenta.

In contrast with the multiring arrangements of electrons in circular semiconductor
quantum dots, we found that the graphene REMs exhibited in all instances a single
$(0,N)$ polygonal-ring molecular structure. Disruptions in the zig-zag
boundary condition along the circular edge behave effectively as crystal-field
effects that {\it pin\/} the electron molecule, yielding single-particle 
densities with broken rotational symmetry that portray directly the azimuthal 
localization of the edge electrons.

\begin{acknowledgments}
This work was supported by the US D.O.E. (Grant No. FG05-
86ER45234).
\end{acknowledgments}

%****************************** begin figure 12 **************************
\begin{figure}[t]
%\centering{\includegraphics[width=8.0cm]{all-exd-ovl-a0-n5-n8_v2.eps}}
\centering{\includegraphics[width=8.0cm]{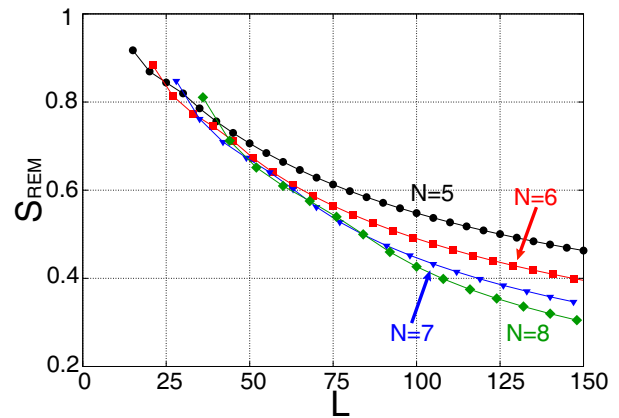}}
\caption{
Overlaps of the parameter-free trial REM states [i.e., for $\alpha=1$, see 
Eq.\ (\ref{remwf})] with the EXD ones, as a function of total angular momentum 
$L$.
}
\label{ovla0_n5_n8}
\end{figure}
%****************************** end figure 12 **************************

\appendix

\section{More on edge states}

The general solution of the eigenvalue equation corresponding to the 
linearized tight-binding Hamiltonian (\ref{tbham}) is of the form 
\begin{eqnarray}
\left(
\begin{array}{c}
\psi_l^{As}
\crcr
\psi_l^{Bs}
\end{array}
\right)
=
\left(
\begin{array}{c}
\chi^{As}_l(r) e^{i[l+(1-s)/2]\phi}
\crcr
\chi^{Bs}_l(r) e^{i[l+(1+s)/2]\phi}
\end{array}
\right),
\label{edge4}
\end{eqnarray}
where $s=\pm$; obviously $s=\pm1$ when occurring in a phase. As a result, the
matrix eigenvalue problem is equivalent to the following set of
equations involving the vector components:
\begin{eqnarray}
-i \hbar v_F \partial_r \chi_l^{B+}-i \hbar v_F (l+1) \frac{\chi_l^{B+}}{r} &=& 
E \chi_l^{A+} \nonumber \\
-i \hbar v_F \partial_r \chi_l^{A+}+i \hbar v_F l \frac{\chi_l^{A+}}{r} &=& 
E \chi_l^{B+},
\label{system}
\end{eqnarray}
where we considered only the case for $s=+$ (the $s=-$ case can be treated in a
similar way).

We are interested in solutions with $E=0$ (the socalled midgap solutions), in 
which case the set of equations (\ref{system}) reduces to
\begin{eqnarray}
\partial_r \chi_l^{B+} + \frac{(l+1)}{r} \chi_l^{B+} &=& 0 \nonumber \\
\partial_r \chi_l^{A+} - \frac{l}{r} \chi_l^{A+} &=& 0.
\label{system2}
\end{eqnarray}
The solutions of these equations are
\begin{eqnarray}
\chi_l^{B+}(r)&=&\chi_l^{B+}(R) \left( \frac{r}{R} \right)^{-l-1} \nonumber \\
\chi_l^{A+}(r)&=&\chi_l^{A+}(R) \left( \frac{r}{R} \right)^l. 
\label{solution}
\end{eqnarray}

The boundary condition is that of a zigzag graphene edge that ends always
on a site of the same lattice, i.e., the condition
\begin{equation}
\chi_l^{B+}(R) =0,
\label{cond}
\end{equation} 
forces the $B+$ component to vanish everywhere on the $B$ sublattice,
yielding the final form
\begin{eqnarray}
\left( \begin{array}{c}
\psi_l^{A+}  \\
\psi_l^{B+}  \\
\end{array} \right)
=
\left( \begin{array}{c}
\chi_l^{A+}(R) \left( \frac{r}{R} \right)^l e^{il\phi}\\
0
\end{array} \right).
\label{solution2}
\end{eqnarray} 

The normalization constant $\chi_l^{A+}(R)$ is easily calculated and was
given in Eq.\ (\ref{edge1}). 

\section{The parameter-free REM ($\alpha=1$)}

As mentioned in Section \ref{vREMcom}, the overlaps ${\cal S}_{\text{REM}} 
\equiv \langle \Phi^{\text{EXD}}_{N,L} | \Phi^{\text{REM}}_{N,L} \rangle $
between the parameter-free REM waves function in the gd-LLL and the EXD ones 
behave in an unsatisfactory way, i.e., they decrease as $L$ increases. The
precise behavior of ${\cal S}_{\text{REM}}$ is displayed in Fig.\ 
\ref{ovla0_n5_n8}, and it contrasts with that of the variational REM,
${\cal S}_{\text{vREM}}$, displayed in Fig.\ \ref{aa_n5_n8}.

%****************************** begin figure 13 **************************
\begin{figure}[t]
%\centering{\includegraphics[width=8.0cm]{ovl-ex1-alphas-n3_v2.eps}}
\centering{\includegraphics[width=8.0cm]{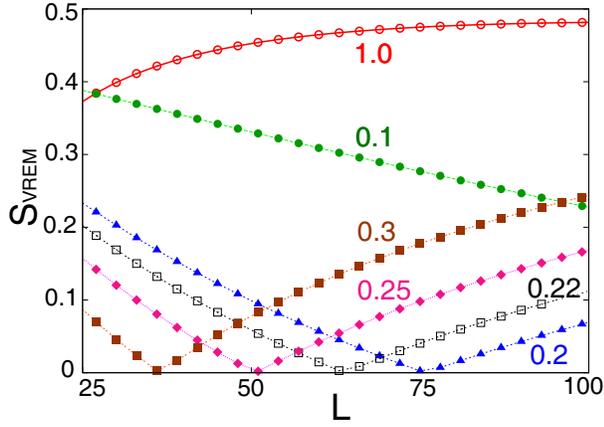}}
\caption{
Overlaps of the variational REM states with the EXD {\it first-excited\/} states
for selected values of the variational parameter $\alpha$ 
[see Eq.\ (\ref{pref2})] as a function of 
total angular momentum $L$. The drops toward zero determine the optimal 
$\alpha$'s for given $L$'s. For $\alpha=1$ (top curve), no such drop to zero 
occurs. 
}
\label{ovlaa1_n5_n8}
\end{figure}
%****************************** end figure 13 **************************

The degradation of the ${\cal S}_{\text{REM}}$ reflects the fact that 
progressively the overlap of the REM wave function with the {\it excited\/} 
EXD states increases with increasing $L$. On the other hand, the optimized 
values of $\alpha$ correspond to variational REM trial functions that have 
practically zero overlap with these excited EXD ones. We have found that such
optimal $\alpha$ values can be found for all studied values of $N$ and $L$.
This is illustrated in Fig.\ \ref{ovlaa1_n5_n8}, where the overlaps of the 
vREM with the first excited EXD state dip toward zero at the optimal 
$\alpha$ values.

\section{Single-particle density} 

We give here the expression for calculating the single-particle density 
$\rho(z)$ for a single many body state $\Phi_L [z]= \sum_I C^I_L \Psi^I_L [z]$, 
where the basis wave functions $\Psi^I_L$ are the Slater determinants defined by
Eq.\ (\ref{detexd}).

Specifically, one has
\begin{eqnarray}
\rho(z) &=&
\langle \Phi_L \vert 
\sum_{k=1}^{N} \delta(z-z_k) 
\vert \Phi_L \rangle \nonumber \\
&=&\sum\limits_{I}\sum\limits_{J} C_L^{I*}C_L^J
\langle \Psi_L^I \vert 
\sum\limits_{k=1}^{N} \delta(z-z_k) 
\vert \Psi_L^J \rangle \nonumber \\
&=&\sum\limits_{I} \vert C^{I}_L\vert^2
\langle \Psi^I_L \vert 
\sum\limits_{k=1}^{N} \delta(z-z_k) 
\vert \Psi^I_L \rangle  \nonumber \\
&=&\sum\limits_I \vert C^{I}_L\vert^2
\sum\limits_{k=1}^{N} 
\psi^*_{l_k^I}(z) \psi_{l_k^I}(z) \nonumber \\
&=& \sum\limits_{I} \vert C^{I}_L\vert^2 
\sum\limits_{k=1}^{N} \frac{l_{k}^{I}+1}{\pi R^2} 
\left( \frac{r}{R} \right)^{2l_k^I},
\label{spden}
\end{eqnarray}
where the edge states $\psi$'s are given by Eq.\ (\ref{edge3}) and 
$l_k^I$ denotes the single-particle angular momenta associated with the
Slater determinant $\Psi^I_L$; naturally $L=\sum_{k=1}^N l^I_k$. The 
single-particle density operator connects in principle Slater determinants
that differ at most in one orbital. However, in the LLL, the conservation
of the total angular momentum implies that there is no pair of 
Slater determinants in the linear superposition of $\Phi_L$ that 
differ precisely by one single orbital; thus one sets $J=I$ when deriving
Eq.\ (\ref{spden}).

\section{Single particle density 
for a superposition of two wave functions}

In Section \ref{pinem} we discussed how the disruptions in the zigzag 
boundary conditions create crystal-field effects that pin the rotating
electron molecule. The effect of this pinning is described through the
linear superposition of two many-body wave functions (EXD and/or REM) with
magic good total angular momenta $L$ and $M$; namely, through a wave function
$\Phi^\text{PIN}$ such that
\begin{equation}
\Phi^\text{PIN} = \frac{1}{\sqrt{2}}
\left( \Phi_L \pm \Phi_M \right),
\label{pinsum}
\end{equation}
where we have dropped the subscript $N$ and superscripts 'EXD' or 'REM' from
the $\Phi_{L(M)}$'s on the r.h.s. 

The $\Phi_{L(M)}$'s are known through their expansions over Slater 
determinants [see Section \ref{secexact}], i.e,
\begin{equation}
 \Phi_L = \sum_I C_L^I \Psi_L^I \qquad\mbox{and}\qquad
 \Phi_M = \sum_J C_M^J \Psi_M^J,
\end{equation}
and the Slater determinants $\Psi_L^I$ and $\Psi_M^J$ are built out of 
single particle states having individual angular momenta $l^I_i$ and
$m^J_j$ such that $\sum_{k=1}^N l^I_k=L$ and  $\sum_{k=1}^N m^J_k=M$.

Using the operator $\hat\rho$ defined by the first line of Eq.\ 
(\ref{spden}), the single-particle density is given by
\begin{eqnarray}
\rho^{\text{PIN}}(z) &=&\langle \Phi^{\text{PIN}} \vert\hat{\rho} 
\vert \Phi^{\text{PIN}} \rangle \nonumber \\
  &=& \frac{1}{2} 
\left( \langle \Phi_L \vert \hat{\rho} \vert \Phi_L \rangle +
\langle\Phi_M\vert \hat{\rho} \vert\Phi_M\rangle \pm  \right. \nonumber \\
&& \left. \langle\Phi_L\vert \hat{\rho} \vert\Phi_M\rangle \pm
\langle\Phi_M\vert \hat{\rho} \vert\Phi_L\rangle \right).
\label{rhoLM}
\end{eqnarray}
The diagonal terms $\langle \Phi_L \vert \hat{\rho} \vert \Phi_L \rangle$ 
and $\langle \Phi_M \vert \hat{\rho} \vert \Phi_M \rangle$ are given by Eq.\ 
(\ref{spden}). Since $\hat\rho$ is a one-body
operator, the cross terms connect Slater determinants that differ precisely by 
one of orbital;\cite{szabobook} we denote by $l^I_p$ and $m^J_q$ the
corresponding pair of indices. By applying the Slater rules described in 
Ref.\ \onlinecite{szabobook} (including bringing the two determinants
into ``maximum coincidence''), one finds:

\begin{widetext}
\begin{eqnarray}
\rho^{\text{PIN}}(z)&=&
  \sum\limits_I\vert C^I_L \vert^2 
  \sum\limits_{k=1}^N
  \frac{(l^I_k+1)}{\pi R^2} \left( \frac{r}{R} \right) ^{2l^I_k} 
+  \sum\limits_J\vert C^J_M \vert^2 
   \sum\limits_{k=1}^{N}
   \frac{( m^J_k+1)}{\pi R^2} \left( \frac{r}{R} \right)^{2m^J_k} \nonumber\\
&\pm&
  \sum\limits_{I,J} {C^I_L}^{*} C^J_M \sigma(L,I,p;M,J,q)
  \frac{ \sqrt{(l^I_p+1)(m^J_q+1)} } {\pi R^2} 
  \left( \frac{r}{R} \right)^{l^I_p+m^J_q} e^{-i\phi(l^I_p-m^J_q)} \\
&\pm&
  \sum\limits_{I,J} {C^I_M}^{*} C^J_L \sigma(M,I,q;L,J,p)
  \frac{ \sqrt{(m^I_q+1)(l^J_p+1)} } {\pi R^2} 
  \left( \frac{r}{R} \right)^{m^I_q+l^J_p} e^{-i\phi(m^I_q-l^J_p)}, \nonumber
\end{eqnarray}
\end{widetext}
where $\sigma(L,I,p;M,J,q)=\pm 1$ depending on the even or odd number of
exchanges of two rows (or columns) needed to bring the two determinants into 
maximum coincidence.

\section{Two-particle conditional probability distribution}
For the conditional probability density [see Eq.\ (\ref{cpdexd})], one has
\begin{eqnarray}
P(z,z_0)&=& 
\sum\limits_I\sum\limits_J C^{*I}_L C^J_L
\langle\Psi^I_L\vert \hat{T} \vert\Psi^J_L\rangle
\label{cpdexd2}
\end{eqnarray}
where the operator $\hat{T}$ is symmetrized; 
\begin{equation}
\hat T=\sum\limits_{i < j}\delta(z-z_i)\delta(z_0-z_j)+
\delta(z-z_j)\delta(z_0-z_i).
\end{equation} 

The matrix elements of $\hat{T}$ between the two Slater determinants 
$\Psi^I_L$ and $\Psi^J_L$ are calculated according to the Slater rules 
for a two-body operator.\cite{szabobook}

\section{More on composite fermions}

There is no reason to {\it a priori\/} restrict the Slater determinants
$\Psi^{\text{IPM}}_{L^*}$ to a certain form. \cite{jain95} Following 
Ref.\ \onlinecite{jain95}, we will restrict the non-interacting $L^*$ 
to the range $-L_0 \leq L^* \leq L_0$, and we will assume that the Slater 
determinants $\Psi^{\text{IPM}}_{L^*}$
are the so-called compact ones. Let $N_n$ denote the number of electrons in the
$n$th Landau Level (LL) with $\sum_{n=0}^t N_n = N$; $t$ is the index
of the highest occupied LL and all the lower LL's with $n \leq t$ are assumed 
to be occupied.  The compact determinants are defined as those in 
which the $N_n$ electrons occupy contiguously the single-particle orbitals 
[$\psi^{\text{DF}}_{p,l}(z)$] of each $n$th LL  $[p+(|l|-l)/2 = n]$ with the 
lowest angular momenta, $l=-n, -n+1, ..., -n+N_n-1$. The compact Slater 
determinants are usually denoted as $[N_0, N_1, ..., N_t]$, and the 
corresponding total angular momenta are given by 
$L^* =(1/2) \sum_{s=0}^t N_s(N_s-2s-1)$.

For the CF theory, the magic angular momenta can be determined by Eq.\ 
(\ref{ltol*}), if one knows the non-interacting $L^*$'s. For $N=6$,
the CF magic $L$'s in any interval $1/(2m-1) \geq \nu \geq 1/(2m+1)$ 
$[15(2m-1) \leq L \leq 15(2m+1)]$, $m=1,2,3,4,...$,  can be found by adding 
$2mL_0=30m$ units of angular momentum to each of the $L^*$'s. 
To obtain the non-interacting $L^*$'s, one needs first to 
construct\cite{jain95} the compact Slater determinants. The compact 
determinants and the corresponding non-interacting $L^*$'s are listed in 
Table \ref{cftapp}. 

%*************************** table II *********************
\begin{table}[t]
\caption{\label{cftapp} Compact non-interacting Slater determinants 
and associated angular momenta $L^*$ for $N=6$ electrons according to
the CF presciption. Both $L^*=-3$ and $L^*=3$ are associated with two compact
states each, the one with lowest energy being the preferred one.}
\begin{ruledtabular}
\begin{tabular}{cr}
 Compact state $[N_0, N_1, ..., N_t]$  &   $L^*$   \\ \hline
$[$1,1,1,1,1,1$]$    &   $-$15   \\
$[$2,1,1,1,1$]$      &   $-$9    \\
$[$2,2,1,1$]$        &   $-$5    \\
$[$3,1,1,1$]$        &   $-$3    \\
$[$2,2,2$]$          &   $-$3    \\
$[$3,2,1$]$          &   0       \\
$[$4,1,1$]$          &   3       \\
$[$3,3$]$            &   3       \\
$[$4,2$]$            &   5       \\
$[$5,1$]$            &   9       \\
$[$6$]$              &   15      \\
\end{tabular}
\end{ruledtabular}
\end{table}
%*************************** end table II *********************

There are nine different values of $L^*$'s, and thus the CF theory for $N=6$ 
predicts that there are always nine magic numbers in any interval 
$15(2m-1) \leq L \leq 15(2m+1)$ between two consecutive JL angular momenta 
$15(2m-1)$ and $15(2m+1)$, $m=1,2,3,...$, For example, using 
Table \ref{cftapp} and Eq.\ (\ref{ltol*}), the CF magic numbers for $N=6$ in the 
interval $15 \leq L < 45$ ($m=1$) are found to be the following nine:
\begin{equation}
15,\;21,\;25,\;27,\;30,\;33,\;35,\;39.
\label{cfmam1}
\end{equation}
In the interval $45 \leq L < 75$ ($m=2$), the CF magic numbers are:
\begin{equation}
45,\;51,\;55,\;57,\;60,\;63,\;65,\;69.
\label{cfmam2}
\end{equation}


\begin{thebibliography}{999}
\bibitem{note2}
There is also a mirror LLL for $l \leq 0$ in analogy with the effect of
reversing the direction of the magnetic field in semiconductor QDs.
\end{thebibliography}
\end{document}